\documentclass[a4paper,10pt,conference]{IEEEtran}
\IEEEoverridecommandlockouts
\usepackage{cite}
\usepackage{amsmath,amssymb,amsfonts,mathtools}
\usepackage{algorithmic}
\usepackage{graphicx}
\usepackage{textcomp}
\usepackage{xcolor}
\usepackage{hyperref}
\usepackage{soul}
\usepackage[amsmath,thmmarks]{ntheorem}
\usepackage{algorithm}
\usepackage{algorithmic}
\usepackage{subcaption}

\usepackage[left=1.62cm,right=1.62cm,top=1.85cm,bottom=4.35cm]{geometry}
\theoremstyle{plain}

\def\BibTeX{{\rm B\kern-.05em{\sc i\kern-.025em b}\kern-.08em
    T\kern-.1667em\lower.7ex\hbox{E}\kern-.125emX}}
\begin{document}

\title{Optimal Pilot Pattern Design for LMMSE Channel Estimation in OFDM Systems with Finite Block Size over Doubly Dispersive Channels
\thanks{The work is partially supported by the National Natural Science Foundation of China (62571467, 62201162).}
}

\author{Xuyao Yu$^{\star}$, Zijun Gong$^{\star,*,\dagger}$, Zhilu Lai$^{\star,\dagger}$ \\
$^{\star}$The Hong Kong University of Science and Technology (Guangzhou), China\\
$^*$HKUST Fok Ying Tung Research Institute, Guangdong, China\\
$^{\dagger}$The Hong Kong University of Science and Technology, Hong Kong}


\maketitle

\begin{abstract}
Pilot pattern design over doubly dispersive channels has regained significant research interest, driven by emerging high-mobility applications in 5G-Advanced and 6G systems, as well as recent developments in Orthogonal Time Frequency Space (OTFS) modulation. This paper addresses the design of LMMSE-optimal pilot patterns for OFDM systems over doubly dispersive channels with finite time-frequency grids. We formulate the problem as A-optimal sensor selection and propose two heuristic algorithms, both combining an initialization stage with local swap refinement. The first employs convex relaxation with randomized rounding, while the second uses greedy selection. Simulations on practical resource block dimensions demonstrate that the proposed designs consistently outperform conventional rectangular and diamond lattice patterns.
\end{abstract}

\begin{IEEEkeywords}
Doubly Dispersive Channel, Pilot Pattern Design, LMMSE, A-Optimal Design, Sensor Selection
\end{IEEEkeywords}

\section{Introduction}

Orthogonal frequency-division multiplexing (OFDM) remains the cornerstone of contemporary wireless standards and is expected to persist in 6G \cite{TMN20256GInterface}. The proliferation of high-mobility scenarios—such as vehicle-to-everything (V2X) and high-speed rail—has rekindled interest in the robustness of OFDM against doubly dispersive (time- and frequency-selective) channels. While emerging modulation schemes like Orthogonal Time Frequency Space (OTFS) explicitly address delay-Doppler coupling, channel estimation in both OTFS and OFDM fundamentally reduces to the same core problem: estimating a correlated two-dimensional (2D) field from scattered time-frequency observations \cite{srivastavaBayesianLearningAided2021,hePilotPatternDesign2022,shengTimeFrequencyDomainChannel2024}. Consequently, the design of pilot patterns over the time-frequency resource grid is of paramount importance for both current 5G-Advanced and future systems.

Unlike quasi-static channels, where uniform spacing in frequency is known to be optimal \cite{negiPilotToneSelection1998,adireddyOptimalPlacementTraining2002,ohnoCapacityMaximizingMMSEOptimal2004}, doubly dispersive channels exhibit correlation across both dimensions. This renders pilot pattern design an inherently 2D sampling problem. Early investigations into this domain primarily relied on structured \textit{lattice} patterns \cite{vaidyanathanMultirateSystemsFilter1993}, such as rectangular or diamond-shaped configurations \cite{tufvessonPilotAssistedChannel1997,negiPilotToneSelection1998,fernandez-getinogarciaPilotPatternsChannel2000,coleriChannelEstimationTechniques2002,mindongOptimalPilotPlacement2002,liPilotsymbolaidedChannelEstimation2000}. A seminal contribution by Choi et al. \cite{ji-woongchoiOptimumPilotPattern2005} established the optimality of diamond-shaped lattices for linear minimum mean square error (LMMSE) estimation under the assumption of an infinite time-frequency grid. However, practical communication systems operate on finite resource blocks (e.g., a 5G NR slot). In such finite regimes, the infinite grids assumption of \cite{ji-woongchoiOptimumPilotPattern2005} breaks down, and lattice patterns often exhibit edge effects or suboptimal alignment with the block boundaries, leaving significant room for improvement.

Latter efforts to address finite-grid constraints include clustered pilot placements \cite{zhangOptimalTrainingPilot2006,shinEfficientDesignDoubly2007} and branch-and-bound search methods for A-optimal design \cite{wangPilotPatternDesign2023}. While the latter can yield performance gains over standard lattice patterns, the combinatorial nature of pilot selection leads to prohibitive computational complexity, even for moderately sized grids. More recently, data-driven approaches using neural networks have been explored to determine pilot locations \cite{soltaniPilotPatternDesign2020,mashhadiPruningPilotsDeep2021,liuPDCEViTNovelPilot2025}. Although these methods show promise, they typically function as black-box optimizers that lack the interpretability and theoretical guarantees required for standardization and reliable deployment.

In contrast to infinite-grid assumptions analyses or high-complexity global searches, this paper addresses a pressing practical question: \textit{Given a finite time-frequency grid, how can we efficiently design an LMMSE-optimal pilot pattern under the constraint of a fixed pilot budget?}

Recognizing that exhaustive search is intractable, we propose two distinct heuristic algorithms to solve the resulting combinatorial optimization problem. Both methods ultimately employ a local swap refinement procedure to improve solution quality. The first approach obtains an initial solution via convex relaxation and randomized rounding, while the second generates its starting point through greedy selection. These two methods offer comparable estimation performance while exhibiting complementary initialization strategies, providing flexible implementation options for practical finite-grid deployments.


The remainder of this paper is organized as follows. Section~\ref{sec:signalmodel} introduces the OFDM signal model and characterizes the second-order statistics of doubly dispersive channels under the WSSUS assumption. Section~\ref{sec:LMMSE} derives the LMMSE channel estimator and formulates the A-optimal pilot pattern design problem. Section~\ref{sec:optimization} presents the two proposed heuristic algorithms. Numerical results are provided in Section~\ref{sec:NumericalResults}, demonstrating the performance gains over conventional lattice patterns. Finally, Section~\ref{sec:conclusion} concludes the paper.

\section{System Model and Channel Statistics}
\label{sec:signalmodel}
\subsection{Input-output Relationship}

We consider an OFDM system with $M$ subcarriers in the frequency domain and $N$ time slots in the time domain, resulting in a time-frequency resource grid of size $M \times N$. Let $\mathbf{G} \in \mathbb{C}^{M \times N}$ denote the discretized time-frequency channel response matrix, where the element $\mathbf{G}[m,n]$ represents the complex channel gain experienced by the symbol transmitted on the $m$-th subcarrier during the $n$-th OFDM symbol period.

The transmitted symbols are arranged in a matrix $\mathbf{X} \in \mathbb{C}^{M \times N}$, where $\mathbf{X}[m,n]$ denotes the complex symbol placed at the $(m,n)$-th resource element. Under the assumption of negligible inter-carrier and inter-symbol interference--justified by the underspread property of practical doubly dispersive channels and proper pulse design--the received signal matrix $\mathbf{Y} \in \mathbb{C}^{M \times N}$ after OFDM demodulation can be expressed as
\begin{equation}
\mathbf{Y} = \mathbf{G} \odot \mathbf{X} + \mathbf{N},
\label{eq:rx_matrix}
\end{equation}
where $\odot$ denotes the element-wise (Hadamard) product, and $\mathbf{N} \in \mathbb{C}^{M \times N}$ is the additive white Gaussian noise (AWGN) matrix with independent and identically distributed entries $\mathbf{N}[m,n] \sim \mathcal{CN}(0, \sigma_n^2)$.

To facilitate linear channel estimation, we vectorize the matrix formulation by stacking the columns of each matrix into a vector. Let $\mathbf{y} = \text{vec}(\mathbf{Y}) \in \mathbb{C}^{MN \times 1}$, $\mathbf{g} = \text{vec}(\mathbf{G}) \in \mathbb{C}^{MN \times 1}$, $\mathbf{x} = \text{vec}(\mathbf{X}) \in \mathbb{C}^{MN \times 1}$, and $\mathbf{n} = \text{vec}(\mathbf{N}) \in \mathbb{C}^{MN \times 1}$. Define the diagonal matrix $\mathbf{B} = \mathrm{Diag}(\mathbf{x}) \in \mathbb{C}^{MN \times MN}$. The vectorized received signal model is then given by
\begin{equation}
\mathbf{y} = \mathbf{B} \mathbf{g} + \mathbf{n}.
\label{eq:rx_vector}
\end{equation}

\subsection{Pilot and Data}
In this paper, the transmitted symbol vector $\mathbf{x}$ is modeled as the superposition of pilot and data symbols:
\begin{equation}
    \mathbf{x} = \mathbf{x}_p + \mathbf{x}_d,
\end{equation}
where $\mathbf{x}_p$ and $\mathbf{x}_d$ are zero-padded pilot and data vectors, respectively. Let $\mathcal{I}_p$ and $\mathcal{I}_d$ denote the index sets of pilot and data symbols such that $\mathcal{I}_p\cup \mathcal{I}_d = \{0, \dots, MN-1\}$ and $\mathcal{I}_p\cap\mathcal{I}_d=\emptyset$,
$$
\mathbf{x}_p[k] = \begin{cases} 
    \substack{\text{pilot}\\\text{symbol}} & k \in \mathcal{I}_p \\ 
    0 & \text{otherwise} 
\end{cases},\
\mathbf{x}_d[k] = \begin{cases} 
    \substack{\text{data}\\\text{symbol}} & k \in \mathcal{I}_d \\ 
    0 & \text{otherwise} 
\end{cases}.
$$
Accordingly, $\mathbf{B}$ can also be expressed as sum of the data part $\mathbf{B}_d=\mathrm{Diag}(\mathbf{x}_d)$ and pilot part $\mathbf{B}_p=\mathrm{Diag}(\mathbf{x}_p)$. 

The pilot symbols are assumed to have constant modulus $\sigma_p$, while the data symbols are i.i.d. $\mathcal{CN}(0,\sigma_d^2)$. Moreover, we assume that the transmitted OFDM $M\times N$ block is average-power-constrained by unit power 1. Let $\beta$ ($\beta \leqslant 1$) denote the fraction of total power allocated to pilot symbol. With $K = |\mathcal{I}_p|$ pilots, the pilot symbol power is given by
\begin{equation}
    \sigma_p^2 = \frac{N\beta}{K}.
    \label{equ:PilotPower}
\end{equation}

For convenience, we further define binary mask vectors $\mathbf{c}_p$ and $\mathbf{c}_d$ as
$$
\mathbf{c}_d[k] = \begin{cases} 
    1 & \text{if } k \in \mathcal{I}_d \\ 
    0 & \text{otherwise} 
\end{cases}, \quad
\mathbf{c}_p[k] = \begin{cases} 
    1 & \text{if } k \in \mathcal{I}_p \\ 
    0 & \text{otherwise} 
\end{cases},
$$
which indicate the pilot and data positions, respectively.

\subsection{Channel Statistics and Covariance Matrix}

We consider a doubly dispersive channel characterized by the delay-Doppler spreading function $V(\tau,\nu)$. Under the wide-sense stationary uncorrelated scattering (WSSUS) assumption~\cite{belloCharacterizationRandomlyTimeVariant1963}, $V(\tau,\nu)$ is a zero-mean circularly symmetric complex Gaussian random process with correlation function
\begin{equation}
\mathbb{E}\!\left[ V(\tau_1, \nu_1) V^*(\tau_2, \nu_2) \right] = \mathcal{P}(\tau, \nu) \, \delta(\tau_1 - \tau_2) \, \delta(\nu_1 - \nu_2),
\label{eq:wssus_correlation}
\end{equation}
where $\mathcal{P}(\tau,\nu)$ denotes the channel scattering function. For practical underspread channels, $\mathcal{P}(\tau,\nu)$ is compactly supported within the rectangular region $\mathcal{D} \triangleq \left[-\frac{\tau_{\mathcal{D}}}{2}, \frac{\tau_{\mathcal{D}}}{2}\right] \times \left[-\frac{\nu_{\mathcal{D}}}{2}, \frac{\nu_{\mathcal{D}}}{2}\right]$, where $\tau_{\mathcal{D}}$ and $\nu_{\mathcal{D}}$ are the maximum delay and Doppler spreads, respectively. The channel spreading factor is defined as
\begin{equation}
\Delta_{\mathcal{D}} \triangleq \tau_{\mathcal{D}} \nu_{\mathcal{D}},
\end{equation}
and the underspread condition implies $\Delta_{\mathcal{D}} \ll 1$. The spreading factor $\Delta_{\mathcal{D}}$ quantifies the joint time-frequency selectivity of the channel.

Under these assumptions, the vectorized channel $\mathbf{g} = \mathrm{vec}(\mathbf{G}) \in \mathbb{C}^{MN \times 1}$ follows a zero-mean circularly symmetric complex Gaussian (CSCG) distribution:
\begin{equation}
\mathbf{g} \sim \mathcal{CN}\!\left( \mathbf{0}, \mathbf{C}_{\mathbf{g}} \right),
\label{eq:channel_distribution}
\end{equation}
where the covariance matrix $\mathbf{C}_{\mathbf{g}} = \mathbb{E}[\mathbf{g} \mathbf{g}^H] \in \mathbb{C}^{MN \times MN}$ has entries given by the two-dimensional symplectic Fourier transform of the scattering function:
\begin{equation}
\mathbf{C}_{\mathbf{g}}[k, l] = \iint_{\mathcal{D}} \mathcal{P}(\tau, \nu) \, e^{j2\pi \left( (n_1 - n_2) T \nu - (m_1 - m_2) F \tau \right)} \, d\nu \, d\tau,
\label{eq:covariance_sfft}
\end{equation}
with $k = (n_1-1)M + m_1$ and $l = (n_2-1)M + m_2$, corresponding to the time-frequency indices $(m_1, n_1)$ and $(m_2, n_2)$, respectively. Here, $T$ denotes the OFDM symbol duration and $F$ denotes the subcarrier spacing.

Moreover, due to the underspread nature of the channel, $\mathbf{C}_{\mathbf{g}}$ exhibits strong correlation among adjacent time-frequency indices, rendering it effectively low-rank. We therefore adopt the reduced-rank representation
\begin{equation}
\mathbf{C}_{\mathbf{g}}
= \mathbf{U}_{r} \mathbf{\Lambda}_{r} \mathbf{U}_{r}^{H},
\label{equ:ChannelCovMatrixEigDecomposition}
\end{equation}
where $\mathbf{\Lambda}_r \in \mathbb{R}^{r \times r}$ contains the $r$ dominant eigenvalues and $\mathbf{U}_r \in \mathbb{C}^{MN \times r}$ comprises the corresponding eigenvectors. Here, $r$ denotes the \emph{effective channel rank}. This low-dimensional characterization will be instrumental for subsequent channel estimation and pilot design.


\section{LMMSE-Optimal Pilot Pattern}
\label{sec:LMMSE}

\subsection{LMMSE Channel Estimation}
We now derive the LMMSE estimate of the channel coefficients, along with the corresponding estimation error covariance. We consider a scenario where the receiver has knowledge of the pilot symbols and the autocovariance of the data symbols. 

The LMMSE estimate of channel coefficients is given as:
\begin{equation}
    \hat{\mathbf{g}} = \mathbf{C}_{\mathbf{g}\mathbf{y}} \mathbf{C}_{\mathbf{y}}^{-1} \mathbf{y},
    \label{equ:LMNMSEFullModelPartialKnown}
\end{equation}
where the involved covariance matrices are expressed as
\begin{subequations}
    \begin{align}
        \mathbf{C}_{\mathbf{g}\mathbf{y}} =& \mathbf{C}_{\mathbf{g} } \mathbf{B}_p^H,  \\
        \mathbf{C}_{\mathbf{y}} =& \mathbf{B}_p \mathbf{C}_{\mathbf{g}} \mathbf{B}_p^H + \sigma_{d}^2 \mathrm{Diag}(\mathbf{c}_d) \odot \mathbf{C}_{\mathbf{g}} + \mathbf{C}_{\mathbf{n}}.
    \end{align}
\end{subequations}
Here, $\mathbf{C}_{\mathbf{n}}=\sigma_n^2\mathbf{I}$ is the covariance matrix of noise $\mathbf{n}$.

The estimation error is defined as $\mathbf{e} = \mathbf{g} - \hat{\mathbf{g}}$. Its covariance matrix, which fully characterizes the estimation performance, is given by:
\begin{equation}
    \begin{aligned}
    \mathbf{C}_{\mathbf{e}} &= \mathbf{C}_{\mathbf{g}} - \mathbf{C}_{\mathbf{g}} \mathbf{B}_p^H \left(\mathbf{B}_p \mathbf{C}_{\mathbf{g}} \mathbf{B}_p^H + \mathbf{C}_{\mathbf{n}} \right)^{-1} \mathbf{B}_p \mathbf{C}_{\mathbf{g}}.
    \end{aligned}
    \label{equ:ErrorCovarianceMatrixSimplify}
\end{equation}
It is evident from \eqref{equ:ErrorCovarianceMatrixSimplify} that the LMMSE estimation error covariance is jointly determined by the channel second-order statistics $\mathbf{C}_{\mathbf g}$ and the pilot pattern encoded in $\mathbf{B}_p$. By substituting the reduced-rank representation in \eqref{equ:ChannelCovMatrixEigDecomposition} into \eqref{equ:ErrorCovarianceMatrixSimplify}, the error covariance matrix can be rewritten as
\begin{equation}
\begin{aligned}
\mathbf{C}_{\mathbf e}
=& \mathbf{U}_r \mathbf{\Lambda}_r \mathbf{U}_r^H - \mathbf{U}_r \mathbf{\Lambda}_r \mathbf{U}_r^H \mathbf{B}_p^H \\
&\left(\mathbf{B}_p \mathbf{U}_r \mathbf{\Lambda}_r \mathbf{U}_r^H \mathbf{B}_p^H + \mathbf{C}_{\mathbf n} \right)^{-1} \mathbf{B}_p \mathbf{U}_r \mathbf{\Lambda}_r \mathbf{U}_r^H .
\end{aligned}
\end{equation}

By applying the matrix inversion lemma, the above expression can be
further simplified as
\begin{equation}
\mathbf{C}_{\mathbf e}
= \mathbf{U}_r \mathbf{A}^{-1} \mathbf{U}_r^H,
\label{equ:ErrorCovariance}
\end{equation}
where the matrix $\mathbf{A}$ is defined as
\begin{equation}
\mathbf{A} = \mathbf{\Lambda}_r^{-1} + \alpha\, \mathbf{U}_r^H \mathrm{Diag}(\mathbf{c}_p) \mathbf{U}_r,
\end{equation}
with $\alpha = \sigma_p^2 / \sigma_n^2 = \beta N / (K \sigma_n^2)$ denoting the pilot signal-to-noise ratio (SNR).
Equivalently, $\mathbf{A}$ can be expressed as
\begin{equation}
\mathbf{A} = \mathbf{\Lambda}_r^{-1} + \alpha \sum_{i=1}^{MN} \mathbf{c}_p[i]\,
\mathbf{u}_i \mathbf{u}_i^H
= \mathbf{\Lambda}_r^{-1} + \alpha \sum_{i\in\mathcal I_p} \mathbf{u}_i \mathbf{u}_i^H,
\label{equ:MatrixADecomposition}
\end{equation}
where $\mathbf{u}_i$ denotes the $i$-th row of $\mathbf{U}_r$. This expression explicitly reveals how the pilot pattern vector $\mathbf{c}_p$ shapes the estimation error covariance through a sum of rank-one updates in the dominant channel subspace.


\subsection{Problem Formulation}
Our objective is to design the pilot pattern $\mathbf{c}_p$ that minimizes the average mean square error (MSE) of the LMMSE estimator. Since $\mathbf{U}_r$ has orthonormal columns, the average MSE is given by $\frac{1}{MN} \mathrm{tr}(\mathbf{C}_{\mathbf{e}}) = \frac{1}{MN} \mathrm{tr}(\mathbf{A}^{-1})$. Minimizing this trace corresponds to the A-optimality criterion in experimental design~\cite{fedorovTheoryOptimalExperiments1972,pukelsheimOptimalDesignExperiments2006,boydConvexOptimization2004}. Accordingly, we formulate the pilot pattern design problem as
\begin{equation}
\begin{aligned}
\underset{\mathbf{c}_p}{\mathrm{min}} & \quad \mathrm{tr}\left( \mathbf{A}^{-1} \right) \\
\mathrm{s.t.} & \quad \mathbf{c}_p \in \{0,1\}^{MN}, \\
& \quad \mathbf{1}_{MN}^{\top} \mathbf{c}_p = K,
\end{aligned}
\label{equ:SensorSelection}
\end{equation}
where $K$ is the prescribed number of pilot symbols.

Problem \eqref{equ:SensorSelection} is a combinatorial optimization problem commonly referred to as a \emph{sensor selection problem} \cite{joshiSensorSelectionConvex2009}. Due to the presence of Boolean selection variables, it is in general NP-hard. Notably, the presence of the prior term $\mathbf{\Lambda}_r^{-1}$ ensures that $\mathbf{A}$ remains invertible even when $K < r$, making the formulation well-defined in the pilot-scarce regime typical of practical systems. 

\section{Proposed Algorithms for A-Optimal Pilot Design}
\label{sec:optimization}

To efficiently obtain high-quality pilot patterns for practical finite grid sizes, we propose two heuristic algorithms. Both follow a two-stage framework: an initialization phase followed by local refinement via Fedorov's exchange method~\cite{fedorovTheoryOptimalExperiments1972}. The two algorithms differ in their initialization strategies--the first employs convex relaxation with randomized rounding, while the second adopts greedy selection. The two approaches achieve comparable MSE performance and offer complementary trade-offs in terms of computational complexity.

\subsection{Convex Relaxation and Rounding}

The first approach relaxes the Boolean constraint $\mathbf{c}_p \in \{0,1\}^{MN}$ to the convex box constraint $\bar{\mathbf{c}}_p \in [0,1]^{MN}$. Introducing an auxiliary variable $\mathbf{Z}$ and applying the Schur complement, the relaxed problem can be cast as a semidefinite program (SDP):

\begin{equation}
\begin{aligned}
\underset{\bar{\mathbf{c}}_p, \mathbf{Z}}{\mathrm{min}} & \quad \mathrm{tr}(\mathbf{Z}) \\
\mathrm{s.t.} & \quad \begin{bmatrix} \mathbf{A} & \mathbf{I} \\ \mathbf{I} & \mathbf{Z} \end{bmatrix} \succeq 0, \\
& \quad \bar{\mathbf{c}}_p \in [0,1]^{MN}, \quad \mathbf{1}_{MN}^{\top} \bar{\mathbf{c}}_p = K,
\end{aligned}
\label{equ:ConvexRelaxationSDPLMI}
\end{equation}

Problem \eqref{equ:ConvexRelaxationSDPLMI} is a standard SDP and can be efficiently solved using interior-point methods via off-the-shelf convex optimization solvers such as self-dual-minimization(SeDuMi). After solving \eqref{equ:ConvexRelaxationSDPLMI}, we obtain an optimal fractional solution $\bar{\mathbf{c}}^{\star}_p \in [0,1]^{MN}$. Although the relaxed solution $\bar{\mathbf{c}}^{\star}_p$ is not directly implementable, it provides valuable insight into how the pilot budget $K$ should be optimally distributed over the time–frequency grid in a fractional sense. \newpage In particular, $\bar{\mathbf{c}}^{\star}_p$ admits a natural probabilistic interpretation as a fractional pilot allocation, where each entry specifies the desired marginal activation probability.

To recover a feasible binary pattern, we employ dependent randomized rounding~\cite{gandhiDependentRoundingIts2006}, whose goal is to generate a random binary vector $\mathbf{c}_p \in {0,1}^{MN}$ satisfying 
$\mathbb{E}\left[\mathbf{c}_{p}[i]\right] = \bar{\mathbf{c}}^{\star}_p[i]$ and $\sum_{i=1}^{MN} \mathbf{c}_{p}[i] = K$.
This procedure preserves the marginal probabilities of the relaxed solution while enforcing the pilot budget constraint almost surely. Algorithm~\ref{alg:DependentRounding} summarizes the dependent randomized rounding procedure.

\begin{algorithm}[t]
\footnotesize
\caption{Dependent Randomized Rounding}
\label{alg:DependentRounding}
\begin{algorithmic}[1]
\REQUIRE Fractional solution $\bar{\mathbf{c}}^{\star}_p \in [0,1]^{MN}$
\ENSURE Binary pilot pattern $\mathbf{c}_p \in \{0,1\}^{MN}$ with $\mathbf{1}_{MN}^{\top}\mathbf{c}_p = K$

\STATE Initialize $\mathbf{c} \leftarrow \bar{\mathbf{c}}^{\star}_p$

\WHILE{there exist indices $i \neq j$ such that $\mathbf{c}[i], \mathbf{c}[j] \in (0,1)$}
    \STATE Select two fractional indices $i$ and $j$
    \STATE Compute\\
        $\delta^{+} \leftarrow \min(1-\mathbf{c}[i]$, $\mathbf{c}[j]), \quad \delta^{-} \leftarrow \min(\mathbf{c}[i],\, 1-\mathbf{c}[j])$
    \STATE Draw a random variable $u \sim \mathrm{Uniform}(0,1)$
    \IF{$u \le \dfrac{\delta^{-}}{\delta^{+} + \delta^{-}}$}
        \STATE $\mathbf{c}[i] \leftarrow \mathbf{c}[i] + \delta^{+}$, $\mathbf{c}[j] \leftarrow \mathbf{c}[j] - \delta^{+}$
    \ELSE
        \STATE $\mathbf{c}[i] \leftarrow c[i] - \delta^{-}$, $\mathbf{c}[j] \leftarrow c[j] + \delta^{-}$
    \ENDIF
\ENDWHILE

\STATE \textbf{return} $\mathbf{c}_p \leftarrow \mathbf{c}$
\end{algorithmic}
\end{algorithm}

\subsection{Greedy Algorithm}

The second approach constructs the pilot set $\mathcal{S}$ greedily by sequentially adding the index that yields the largest reduction in the A-optimal objective. Let $\mathbf{A}_{\mathcal{S}} = \mathbf{\Lambda}_r^{-1} + \alpha \sum_{i \in \mathcal{S}} \mathbf{u}_i \mathbf{u}_i^H$. The marginal gain of adding candidate $j \notin \mathcal{S}$ is
\begin{equation}
\Delta_j = \mathrm{tr}(\mathbf{A}_{\mathcal{S}}^{-1}) - \mathrm{tr}(\mathbf{A}_{\mathcal{S}\cup\{j\}}^{-1}) = \frac{\alpha\,\mathbf{u}_j^H \mathbf{A}_{\mathcal{S}}^{-2} \mathbf{u}_j}{1 + \alpha\,\mathbf{u}_j^H \mathbf{A}_{\mathcal{S}}^{-1} \mathbf{u}_j}.
\label{equ:greedy_gain}
\end{equation}

At each iteration, the index maximizing $\Delta_j$ is selected and $\mathbf{A}_{\mathcal{S}}^{-1}$ is updated via the Sherman-Morrison formula given as
\begin{equation}
\mathbf{A}_{\mathcal{S}\cup\{j\}}^{-1} = \mathbf{A}_{\mathcal{S}}^{-1} - \frac{\alpha\,\mathbf{A}_{\mathcal{S}}^{-1} \mathbf{u}_j \mathbf{u}_j^{H} \mathbf{A}_{\mathcal{S}}^{-1}} {1+\alpha\,\mathbf{u}_j^{H} \mathbf{A}_{\mathcal{S}}^{-1} \mathbf{u}_j}.
\label{eq:greedy_smw}
\end{equation}
The procedure repeats until $K$ pilots are selected. The complexity of this greedy algorithm is $O(K r^2 MN)$, scaling linearly with the grid size. Algorithm~\ref{alg:greedy} summarizes the proposed greedy algorithm.

\begin{algorithm}[t]
\footnotesize 
\caption{Greedy Algorithm}
\label{alg:greedy}
\begin{algorithmic}[1]
\REQUIRE Pilot budget $K$, vectors $\{\mathbf{u}_i\}_{i=1}^{MN}$, parameters $\mathbf{\Lambda}_r$, $\alpha$
\ENSURE Binary pilot pattern $\mathbf{c}_p \in \{0,1\}^{MN}$

\STATE Initialize $\mathcal{S} \leftarrow \emptyset$,
$\mathbf{A}_{\mathcal{S}}^{-1} \leftarrow \mathbf{\Lambda}_r$
\FOR{$k = 1$ \TO $K$}
    \FOR{each $j \notin \mathcal{S}$}
        \STATE Compute marginal gain $\Delta_j$ using \eqref{equ:greedy_gain}
    \ENDFOR
    \STATE Select $j^{\star} = \arg\max_{j \notin \mathcal{S}} \Delta_j$
    \STATE Update $\mathcal{S} \leftarrow \mathcal{S} \cup \{j^{\star}\}$
    \STATE Update $\mathbf{A}_{\mathcal{S}}^{-1}$ using \eqref{eq:greedy_smw}
\ENDFOR
\RETURN Binary $\mathbf{c}_p$ corresponding to $\mathcal{S}$
\end{algorithmic}
\end{algorithm}

\subsection{Local Swap Refinement}

Both initialization methods are followed by a local swap procedure to further improve the solution quality. Starting from any feasible pilot set $\mathcal{S}$ with $|\mathcal{S}| = K$, the algorithm iteratively swaps a selected index $i \in \mathcal{S}$ with an unselected index $j \notin \mathcal{S}$ if such an exchange strictly reduces $\mathrm{tr}(\mathbf{A}^{-1})$. The procedure terminates when no improving swap exists, yielding a locally optimal pattern. Efficient evaluation of candidate swaps is enabled by the Sherman-Morrison-Woodbury formula. The proposed local swap method is summarized in Algorithm~\ref{alg:local_swap}.

\begin{algorithm}[t]
\footnotesize
\caption{Local Swap Improvement (Fedorov's Exchange)}
\label{alg:local_swap}
\begin{algorithmic}[1]
\REQUIRE Vectors $\{\mathbf{u}_i\}_{i=1}^{MN}$, prior matrix $\mathbf{\Lambda}_r$, parameter $\alpha$
\STATE Initialize any pilot index set $\mathcal{S}\subseteq\{1,\dots,MN\}$ with $|\mathcal{S}|=K$
\STATE $\mathbf{A}_{\mathcal{S}} \leftarrow \mathbf{\Lambda}_r^{-1} + \alpha \sum_{i\in\mathcal{S}} \mathbf{u}_i\mathbf{u}_i^H$
\WHILE{$\exists\, i\in\mathcal{S},\, j\notin\mathcal{S}$ such that $\mathrm{tr}\!\left(\mathbf{A}_{\mathcal{S}\setminus\{i\}\cup\{j\}}^{-1}\right) < \mathrm{tr}\!\left(\mathbf{A}_{\mathcal{S}}^{-1}\right)$}
    \STATE $\mathcal{S} \leftarrow \mathcal{S}\setminus\{i\}\cup\{j\}$
    \STATE $\mathbf{A}_{\mathcal{S}} \leftarrow \mathbf{A}_{\mathcal{S}} - \alpha\mathbf{u}_i\mathbf{u}_i^H + \alpha\mathbf{u}_j\mathbf{u}_j^H$
\ENDWHILE
\RETURN $\mathcal{S}$
\end{algorithmic}
\end{algorithm}

\section{Numerical Results}
\label{sec:NumericalResults}
In this section, we evaluate the performance of the proposed pilot pattern design algorithms through numerical simulations. The simulation parameters are set to $M = 12$ subcarriers and $N = 14$ OFDM symbols, corresponding to the dimensions of a single resource block (RB) in practical systems such as 5G NR. The channel is generated according to the WSSUS model described in Section~\ref{sec:signalmodel}, with a separable scattering function following a truncated exponential power delay profile and a Jakes Doppler spectrum. 

\subsection{Visualization of Designed Pilot Patterns}

\begin{figure*}[htbp]
    \centering
    \includegraphics[width=\linewidth]{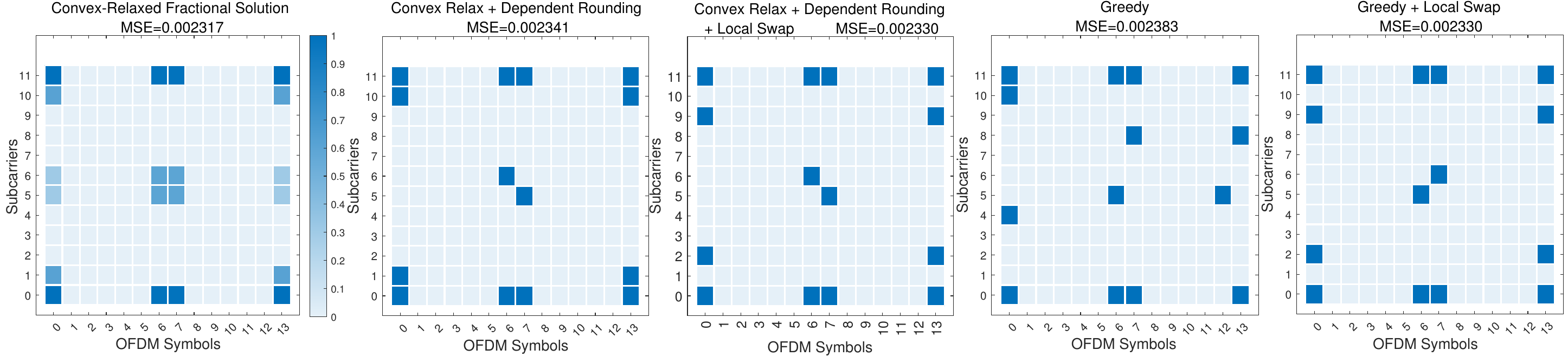}
    \caption{Designed pattern with 8\% of pilot budget.}
    \label{fig:SingleRBExample_8Percent}
    \vspace{-5mm}
\end{figure*}

We first present illustrative examples of the pilot patterns obtained by the proposed algorithms. Fig.~\ref{fig:SingleRBExample_8Percent} shows the designed patterns for $K = 14$ ($8\%$ of the grid), $\Delta_{\mathcal{D}}=0.001$ and SNR$=10$ dB. We display five patterns: (i) the fractional solution from convex relaxation, (ii) the binary pattern after dependent randomized rounding, (iii) the refined pattern after applying local swap to the rounded solution, (iv) the pattern obtained by greedy initialization, and (v) the refined pattern after applying local swap to the greedy solution.

Several observations are in order. First, the convex-relaxed fractional solution exhibits a smooth spatial distribution, concentrating higher weights in regions where the channel correlation structure dictates greater sampling density. Second, both initialization methods yield structured patterns. Third, and most importantly, the local swap refinement consistently reduces the MSE in all cases, as indicated by the MSE values reported beneath each subfigure. Notably, the final MSE values achieved by both algorithms after local swap refinement are nearly identical, confirming that the two initialization strategies converge to solutions of comparable quality.

\subsection{MSE Performance Comparison}
In this subsection, we compare the channel estimation MSE of the proposed algorithms against conventional lattice-based pilot patterns. As benchmarks, we consider the optimal rectangular and diamond-shaped patterns, where the pilot spacings in time and frequency are optimized for each pilot density through exhaustive search.

\begin{figure}[htbp]
    \centering
    \includegraphics[width=\linewidth]{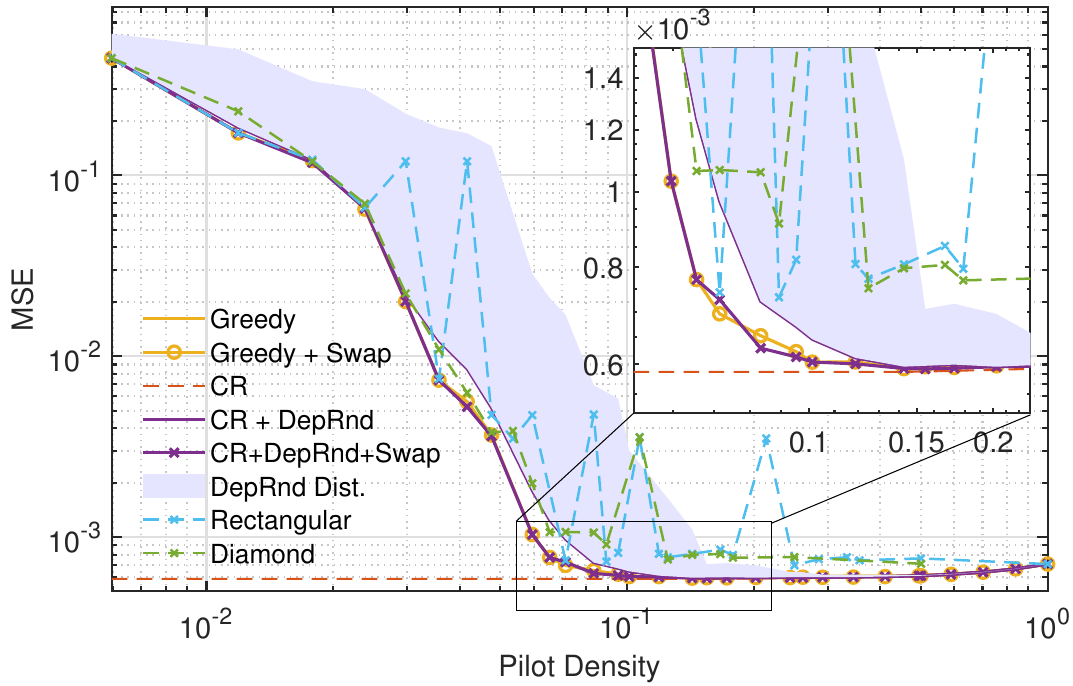}
    \caption{MSE versus pilot density for different pilot patterns. The following abbreviations are used: ``CR'' for Convex Relaxation, ``DepRnd'' Dependent Rounding, ``DepRnd Dist.'' for the distribution of Dependent Rounding results.}
    \label{fig:CompareWithRectDiamond}
    \vspace{-6mm}
\end{figure}

Fig.~\ref{fig:CompareWithRectDiamond} shows the MSE as a function of the pilot density $K / (MN)$ with  $\Delta_{\mathcal{D}}=0.005$ and SNR$=20$ dB. The proposed algorithms consistently outperform the rectangular and diamond shape patterns across all pilot densities. Furthermore, the local swap refinement consistently improves upon the initial solutions obtained from both greedy initialization and convex relaxation with rounding. The improvement is particularly noticeable at moderate pilot densities, where the initial heuristic solutions may deviate from local optima.

\subsection{Impact of Channel Spreading Factor}
\begin{figure}[htbp]
    \centering
    \includegraphics[width=0.75\linewidth]{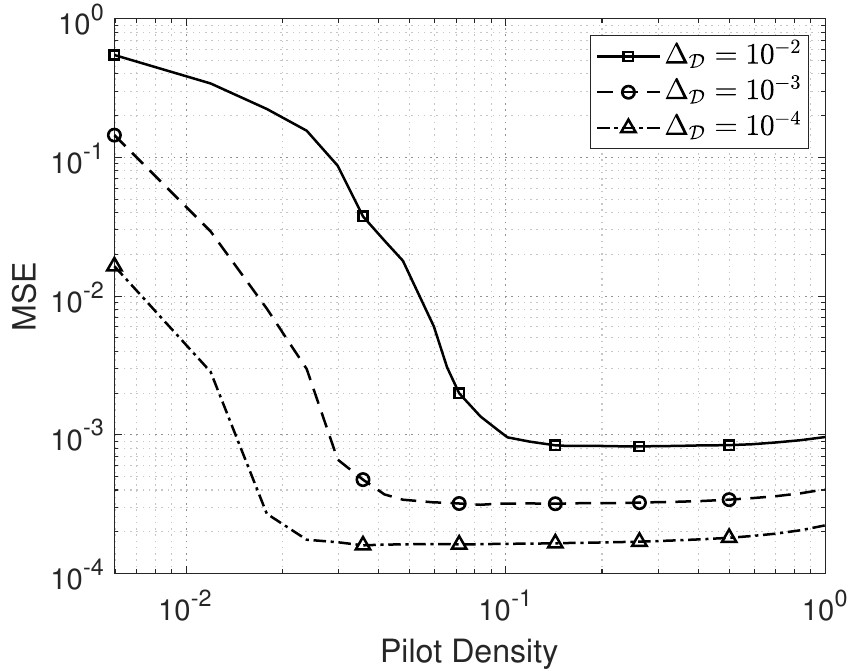}
    \caption{Channel estimation MSE versus pilot density for different channel spreading factors $\Delta_{\mathcal{D}}$.}
    \label{fig:ProposedMethodVaryingSpreadFactor}
    \vspace{-5mm}
\end{figure}

Fig.~\ref{fig:ProposedMethodVaryingSpreadFactor} examines the impact of the channel spreading factor $\Delta_{\mathcal{D}}$ on the estimation performance. We fix the SNR = 20 dB and vary the pilot density for three different channel conditions, corresponding to low, medium, and high mobility scenarios. As expected, the MSE degrades as the channel spreading factor increases, since a larger $\Delta_{\mathcal{D}}$ implies weaker correlation across the time-frequency grid and, consequently, a higher effective channel rank. For each spreading factor, there exists a critical pilot density below which the MSE degrades sharply. This threshold increases with $\Delta_{\mathcal{D}}$, indicating that channels with larger spreading factors demand a higher minimum pilot investment to achieve reliable estimation.

\subsection{Structural Properties of Optimal Patterns}

\begin{figure}[htbp]
    \centering
    \begin{subfigure}[b]{0.99\linewidth}
        \centering
        \includegraphics[width=\linewidth]{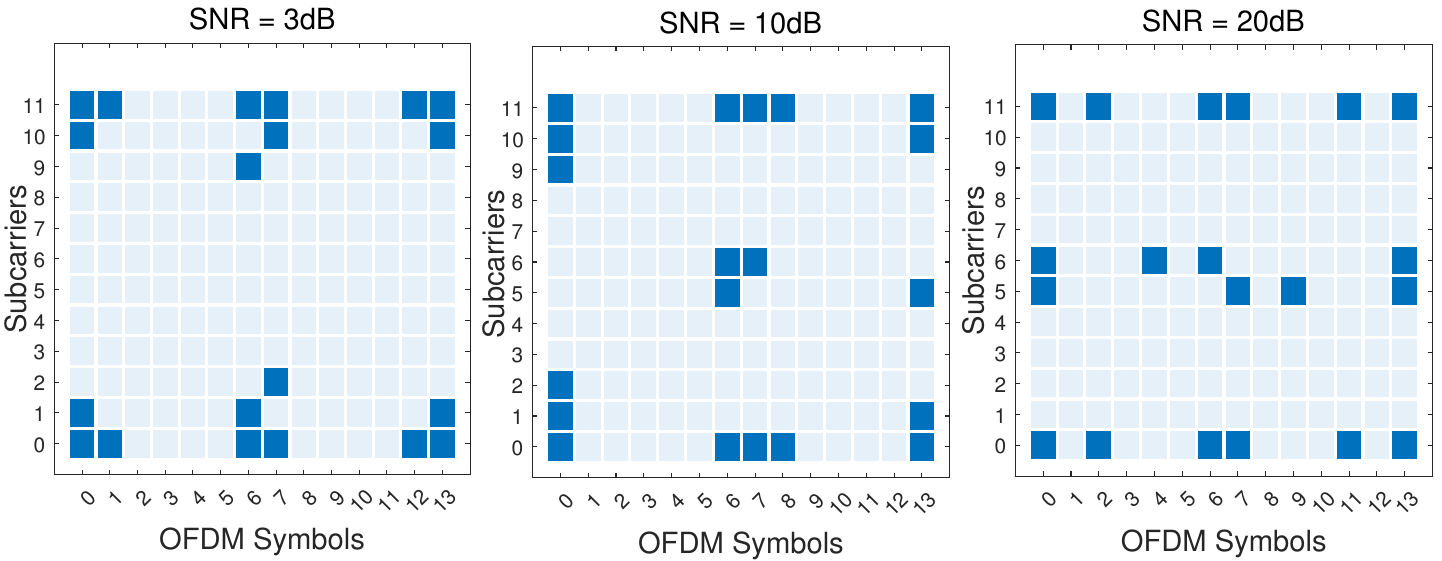}
        \caption{Designed pilot pattern versus symbol SNR.}
        \label{fig:PatternInstance_VaryingSNR}
    \end{subfigure}
    \hfill  
    \begin{subfigure}[b]{0.99\linewidth}
        \centering
        \includegraphics[width=\linewidth]{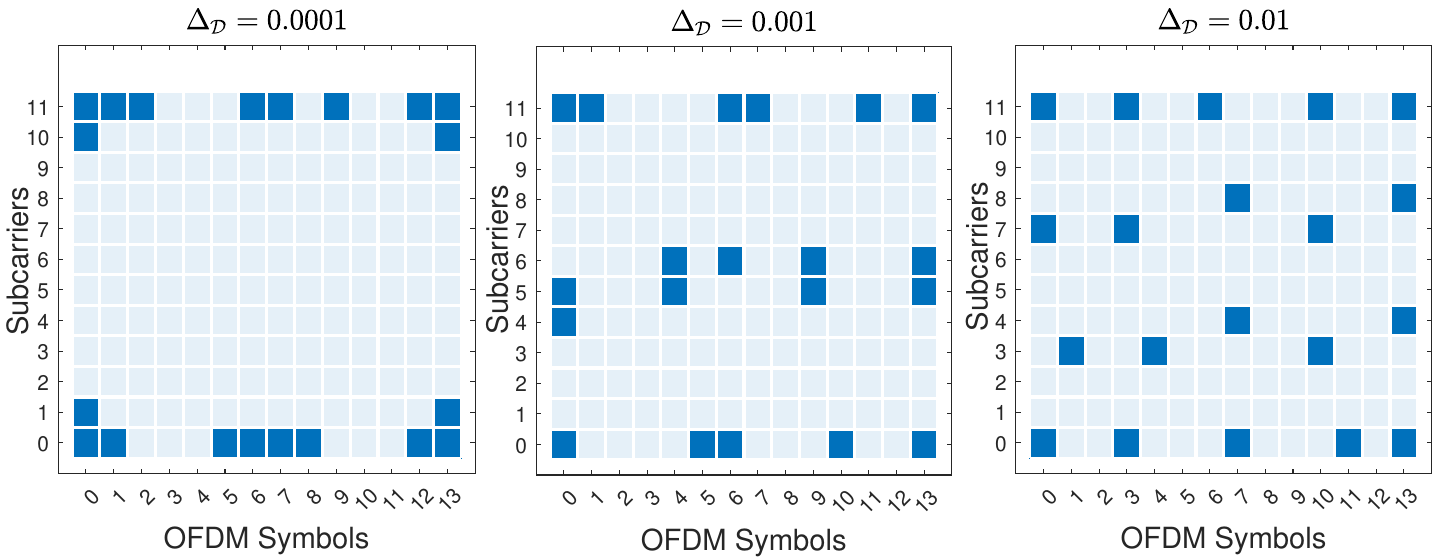}
        \caption{Designed pilot pattern versus channel spreading factor.}  
        \label{fig:PatternInstance_VaryingSpreadFactor}
    \end{subfigure}
    \caption{Designed pattern of proposed method with different SNR and spreading factor.}  
    \vspace{-5mm}
\end{figure}

To gain further insight into the structure of optimal pilot patterns, we examine how the designed patterns vary with SNR and the channel spreading factor. Fig.~\ref{fig:PatternInstance_VaryingSNR} illustrates the optimized patterns for a fixed pilot budget ($K = 20$) and channel spreading factor ($\Delta_{\mathcal{D}} = 0.001$) under three different SNR levels: $3$~dB, $10$~dB, and $20$~dB. A noteworthy trend emerges: at low SNR, the optimal pattern tends to concentrate pilots in clustered formations, whereas at high SNR, the pilots are distributed more uniformly across the time-frequency grid. This behavior can be understood from an information-theoretic perspective. In the low-SNR regime, the observation is dominated by noise, and clustering pilots enhances the effective local SNR, thereby improving the reliability of channel estimates in correlated neighborhoods. Conversely, in the high-SNR regime, estimation accuracy is primarily limited by the channel's degrees of freedom; uniformly spreading the pilots across the grid maximizes the coverage of the dominant channel subspace, leading to better overall reconstruction.

Fig.~\ref{fig:PatternInstance_VaryingSpreadFactor} depicts the optimized patterns for a fixed pilot budget ($K=20$) and SNR = 20 dB under three different channel spreading factors: $\Delta_{\mathcal{D}}=0.0001$ (low mobility), $0.001$ (moderate mobility), and $0.01$ (high mobility). As the spreading factor increases, the channel correlation decays more rapidly in both time and frequency, effectively enlarging the channel's degrees of freedom. Consequently, the optimal pilot pattern transitions from a more clustered arrangement to a more uniformly dispersed configuration, reflecting the need to capture independent channel variations across a broader region of the time-frequency grid.

These structural insights not only validate the effectiveness of the proposed algorithms but also provide interpretable design guidelines for practical system deployment under varying channel conditions and SNR regimes.

\section{Conclusion}
\label{sec:conclusion}
In this paper, we investigated A-optimal pilot pattern design for LMMSE channel estimation in finite OFDM resource blocks over doubly dispersive channels. Two efficient heuristic algorithms were proposed, both employing local swap refinement following distinct initialization strategies. Numerical results show that the proposed designs consistently outperform conventional lattice patterns. Structural analysis reveals intuitive adaptations: clustered pilots at low SNR, uniform distribution at high SNR or large channel spreading, and boundary concentration when the pilot budget approaches the effective channel rank.

\bibliographystyle{IEEEtran}
\bibliography{ref}

\end{document}